\documentclass{appolb}
\usepackage{graphicx}
\begin{document}
% \eqsec  % uncomment this line to get equations numbered by (sec.num)
\title{Rapidity scan in heavy ion collisions at $\sqrt{s_{\rm NN}}=72$~GeV using a viscous hydro + cascade model
%\thanks{Presented at ...}%
% you can use '\\' to break lines
}
\author{Iurii Karpenko
\address{SUBATECH, IN2P3/CNRS, %
  Universit\'{e} de Nantes, IMT Atlantique,\\ %
  4 rue Alfred Kastler, 44307 Nantes cedex 3, %
  France }
}
\maketitle
\begin{abstract}
In this note we discuss the rapidity dependence of the initial and final conditions for hydrodynamic evolution as well as the resulting basic hadronic observables in heavy ion collisions at $\sqrt{s_{\rm NN}}=72$~GeV in the framework of a viscous hydro+cascade model \texttt{vHLLE+UrQMD}. The resulting rapidity dependences are driven to a big extent by the initial state, which is simulated with the UrQMD cascade. The results can serve as a prediction for future experiments such as the \texttt{AFTER@LHC} or the BES-II program at STAR.
\end{abstract}
  
\section{Introduction}
Heavy ion collisions at the energies corresponding to the upper part of the RHIC Beam Energy Scan program represent an interesting case for exploration in the framework of hydrodynamic approach. From one side, the distributions of produced hadrons are approximately boost invariant in the range of a few units of rapidity around midrapidity - the picture well established in heavy ion collisions at the higher collision energies (top RHIC and LHC). From the other side, at such energies the effects of finite baryon density start to play a role, due to finite baryon stopping.

In this respect the AFTER@LHC program \cite{Brodsky:2012vg, Trzeciak:2017csa, Kikola:2017hnp, Massacrier:2017oeq} looks promising, as it plans to use the LHC beam in a fixed target mode operation, colliding heavy ios (lead-lead, lead-xenon) at $\sqrt{s_{\rm NN}}=72$~GeV. Because of the fixed-target nature of the experiment one can measure hadron production in a wide interval of rapidities, up to the projectile/target rapidity. Also, STAR experiment at RHIC plans Phase II of the Beam Energy Scan program with the upgraded detector which will allow for a larger rapidity coverage. In this work we assess a rapidity dependence of the properties of matter created in such reactions, in the framework of viscous hydrodynamic + cascade model, vHLLE+UrQMD. The structure is as follows: in Section 2 we briefly describe the model and show the rapidity dependence of thermodynamic parameters at the end of the hydro stage, in Section 3 we discuss the resulting observables, and make conclusions in Section 4.

\section{Model and thermodynamics at particlization}

\begin{figure}
 \begin{center}
 \includegraphics[width=0.8\textwidth]{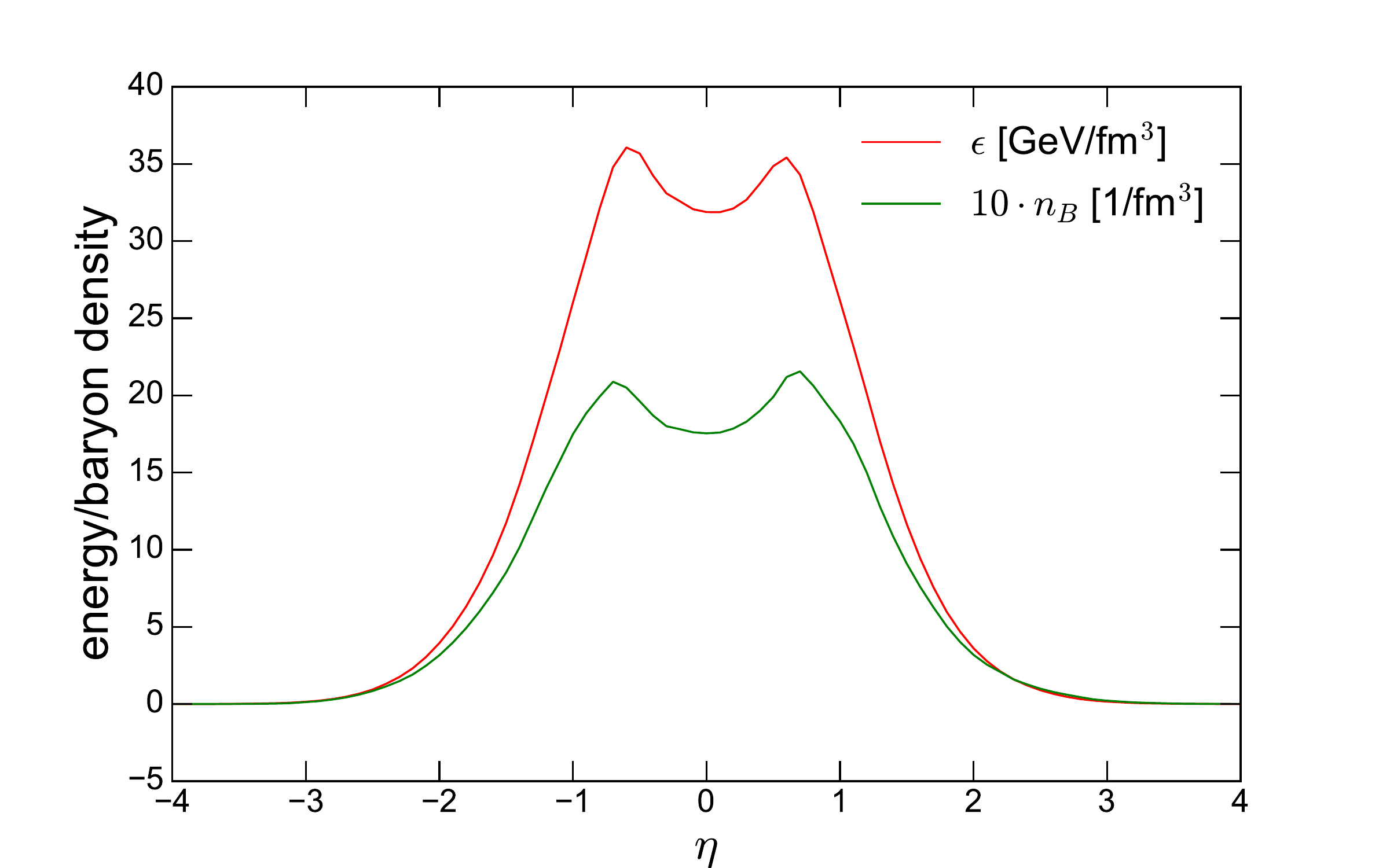}
 \end{center}
 \caption{Event-averaged initial energy and baryon densities for the hydrodynamic stage, in the center of the transverse plane, x=y=0, as a function of the spacetime rapidity $\eta$. The simulation is performed for 0-5\% central PbPb collison at $\sqrt{s_{\rm NN}}=72$~GeV.}\label{fig-densities}
\end{figure}

We use a hydro+cascade model vHLLE+UrQMD to simulate the heavy ion reactions at $\sqrt{s_{\rm NN}}=72$~GeV. The initial stage of the reaction is described with hadron/string cascade UrQMD \cite{Bass:1998ca}. At the hypersurface of constant Bjorken proper time $\tau_0=\sqrt{t^2-z^2}=0.64$~fm/c the system is {\it fluidized}, i.e.~the energy-momenta and charges of initial state particles are distributed to the hydrodynamic grid. This way we obtain a 3 dimensional initial state with nonzero baryon, electric charge and - in event-by-event calculation - strangeness densities, which are propagated in the hydrodynamic stage by solving the corresponding charge conservation equations. On Fig.~\ref{fig-densities} we show event-averaged initial energy and baryon densities in the center of the transverse plane, x=y=0, as a function of the spacetime rapidity $\eta$. We note that the baryon density distribution from the UrQMD initial stage is different from the other dynamical initial state model in \cite{Shen:2017bsr}. However, in our case the initial state also has certain longitudinal flow which is stronger than the boost invariant one. This results in a stronger propagation of the baryon density to higher geometrical rapidities. The initial state also contains a finite angular momentum of the system, which is important for the description of global $\Lambda$ polarization in heavy ion collisions \cite{Karpenko:2016jyx}. The subsequent evolution is treated in the framework of a 3 dimensional event-by-event viscous hydrodynamic approximation with vHLLE code \cite{Karpenko:2013wva}.

When the energy density in the local rest frame of a fluid element reaches the value of $\epsilon_{\rm sw}=0.5$~GeV/fm$^3$, and the equation of state of matter is well described in terms of a non-interacting hadron resonance gas, a so-called particlization is performed, i.~e.~change from fluid medium to particle description. At this point we use the Cooper-Frye prescription \cite{Cooper:1974mv}. The post-hydrodynamic evolution - hadronic rescatterings and resonance decays - is performed again with the UrQMD cascade.

The value of $\epsilon_{\rm sw}=0.5$~GeV/fm$^3$ has been chosen in a previous analysis \cite{Karpenko:2015xea} to optimally describe the observables (including hadron chemistry) in the full range of RHIC BES energies, therefore we use this setting as ``universal'' and use it to check what happens in a ``rapidity scan'' at a fixed collision energy instead of a collision beam energy scan.

\begin{figure}
 \hspace{-25pt}\includegraphics[width=0.6\textwidth]{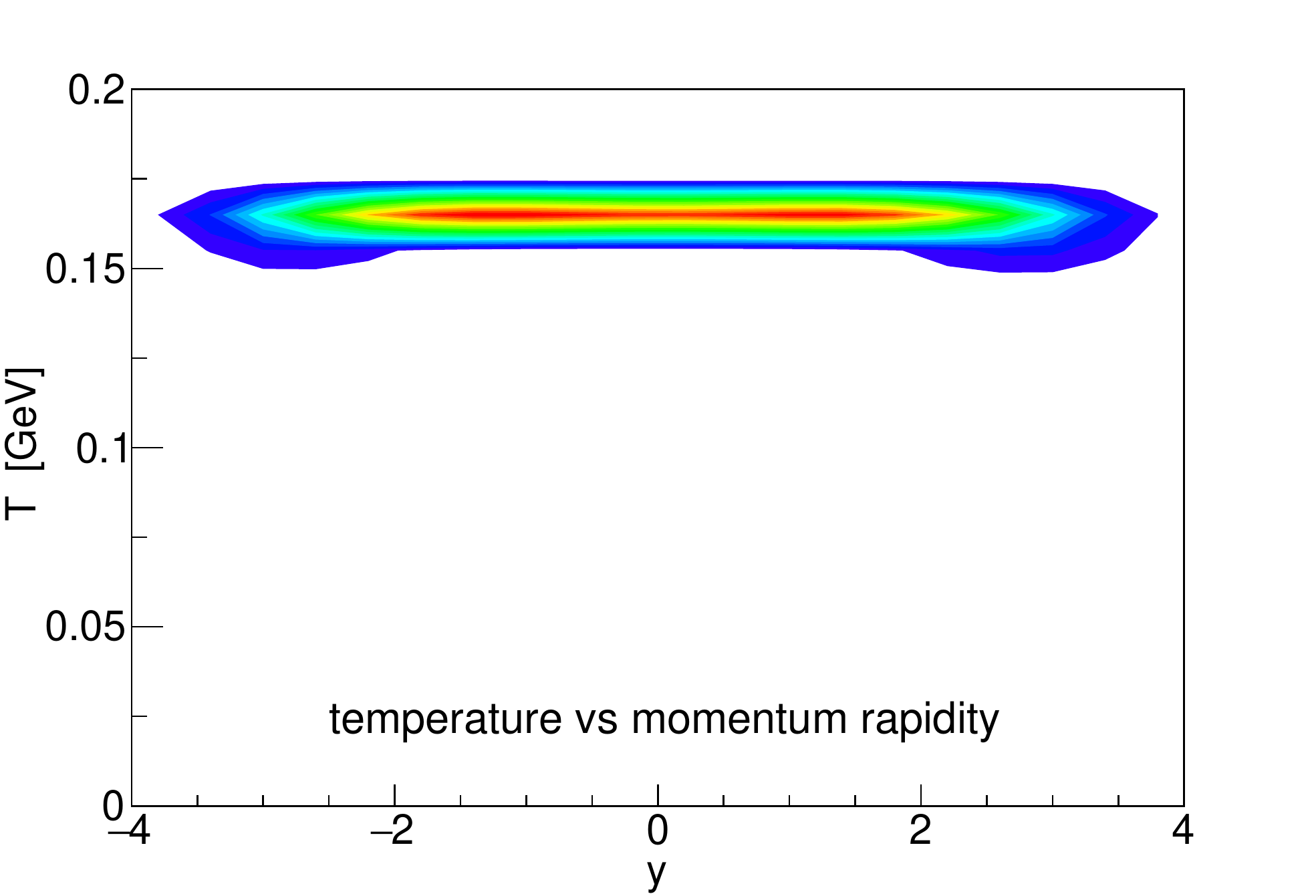}
 \hspace{-15pt}\includegraphics[width=0.6\textwidth]{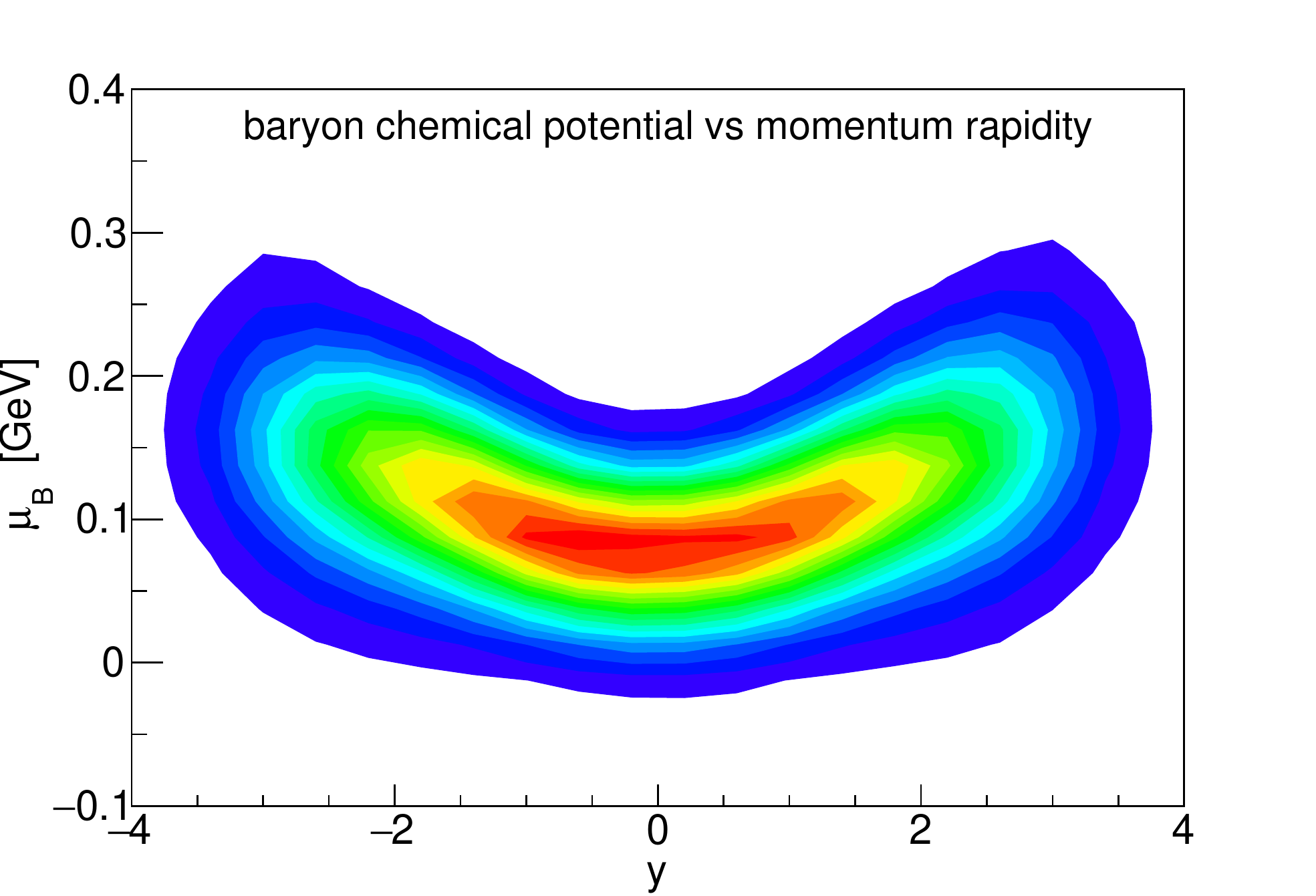}
 \caption{Distribution of temperature and baryon chemical potentials across the particlization hypersurface as a function of momentum rapidity of produced hadrons. Color encode the rate of particles produced with a given value of rapidity from a surface element with given temperature or baryon chemical potential. The event-by-event hydrodynamic calculation is performed for 20-30\% central PbPb collison at $\sqrt{s_{\rm NN}}=72$~GeV.}\label{figT}
\end{figure}

\begin{figure}
 \begin{center}
 \includegraphics[width=0.8\textwidth]{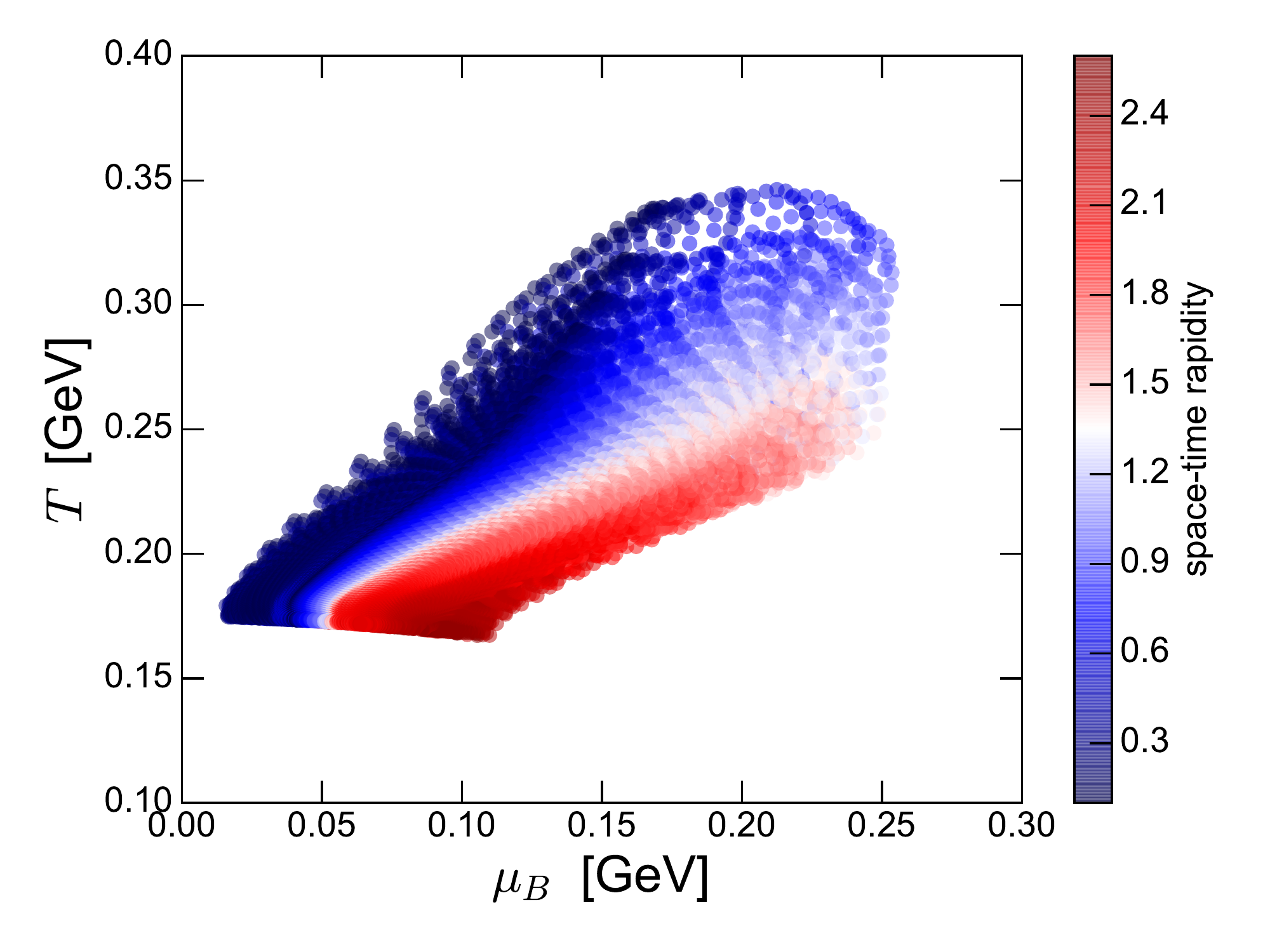}\\
 \vspace{-20pt}
 \end{center}
 \caption{Trajectories (represented by points) on the $T-\mu_{\rm B}$ plane, covered by the slices of the fluid corresponding to different rapidities. Color encode the rapidity of the slice. The simulation is performed for averaged initial state corresponding to 20-30\% central Pb-Pb collisions at $\sqrt{s_{\rm NN}}=72$~GeV.}\label{fig-phaseDiag}
\end{figure}

Whereas the energy density at the particlization surface is fixed by construction, the charge densities (baryon, electric and strange) are given by the hydrodynamic evolution. As a result the densities vary across the hypersurface. The Cooper-Frye prescription used to sample hadrons at the particlization hypersurface is based on grand canonical ensemble. Therefore we recalculate temperature, baryon, electric and strange chemical potentials according to hadron gas EoS for given energy, baryon and strange densities on each element of particlization hypersurface. Naturally, the resulting temperature and chemical potentials vary across the hypersurface as well, and their distribution is visualized on Fig.~\ref{figT}. The color coding on this figure correspond to intensity of particle emission from surface elements with given spacetime rapidity and given value of temperature. The actual range of temperatures and chemical potentials is even wider if one looks at all hypersurface elements irrespective of their contribution to the total particle production. Fig.~\ref{figT} demonstrates that (i) the average baryon chemical potential grows away from midrapidity, (ii) the variation of the baryon chemical potential even at fixed spacetime rapidity is quite substantial, which constrasts the conventional picture of particular collision energy probing via chemical freeze-out a particular point on the QCD phase diagram \cite{Andronic:2017pug}. In fact as one can see from Fig.~\ref{fig-phaseDiag}, due to inhomogenity of the initial energy and baryon densities in all space directions, different fluid elements cover certain areas on the $T-\mu_{\rm B}$ plane in the course of their evolution. Fluid slices with higher spacetime rapidity cover strips at larger $\mu_{\rm B}$ as initial ratio of baryon to energy density grows with (modulus of) rapidity. This way the different rapidity slices scan different areas of the phase diagram in the course of their evolution.

Whereas the trajectories in the $T-\mu_{\rm B}$ plane on Fig.~\ref{fig-phaseDiag} do not depend on the particlization criterion, the temperature and chemical potential distributions on the particlization hypersurface clearly do. However, the main factor here is independent propagation of baryon current in hydrodynamic phase. Therefore a different choice of particlization criterion, e.g.~a fixed temperature, will result in qualitatively similar distribution of baryon chemical potential at the particlization hypersurface.

\section{Rapidity dependence of observables}
\begin{figure}
 \includegraphics[width=0.55\textwidth]{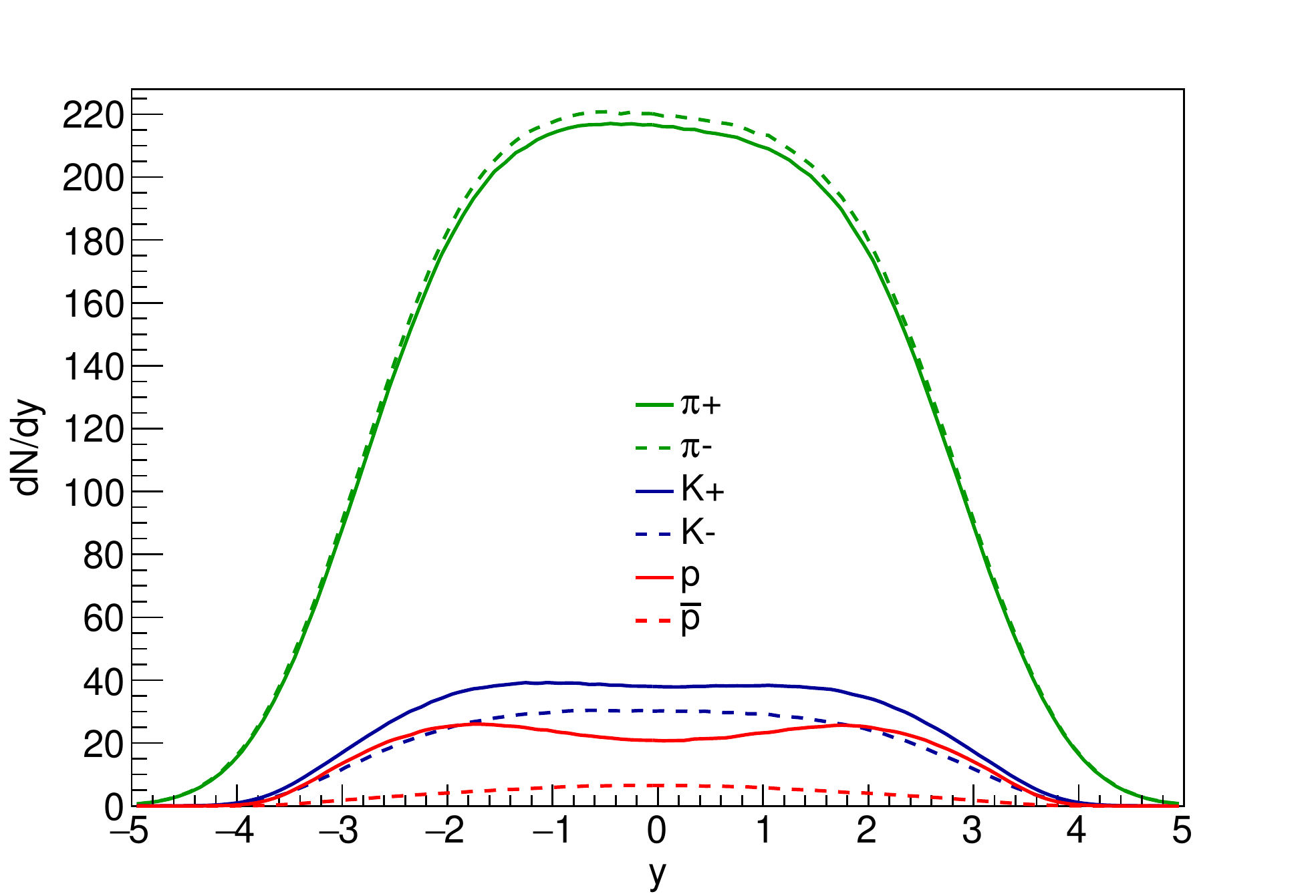}
 \includegraphics[width=0.55\textwidth]{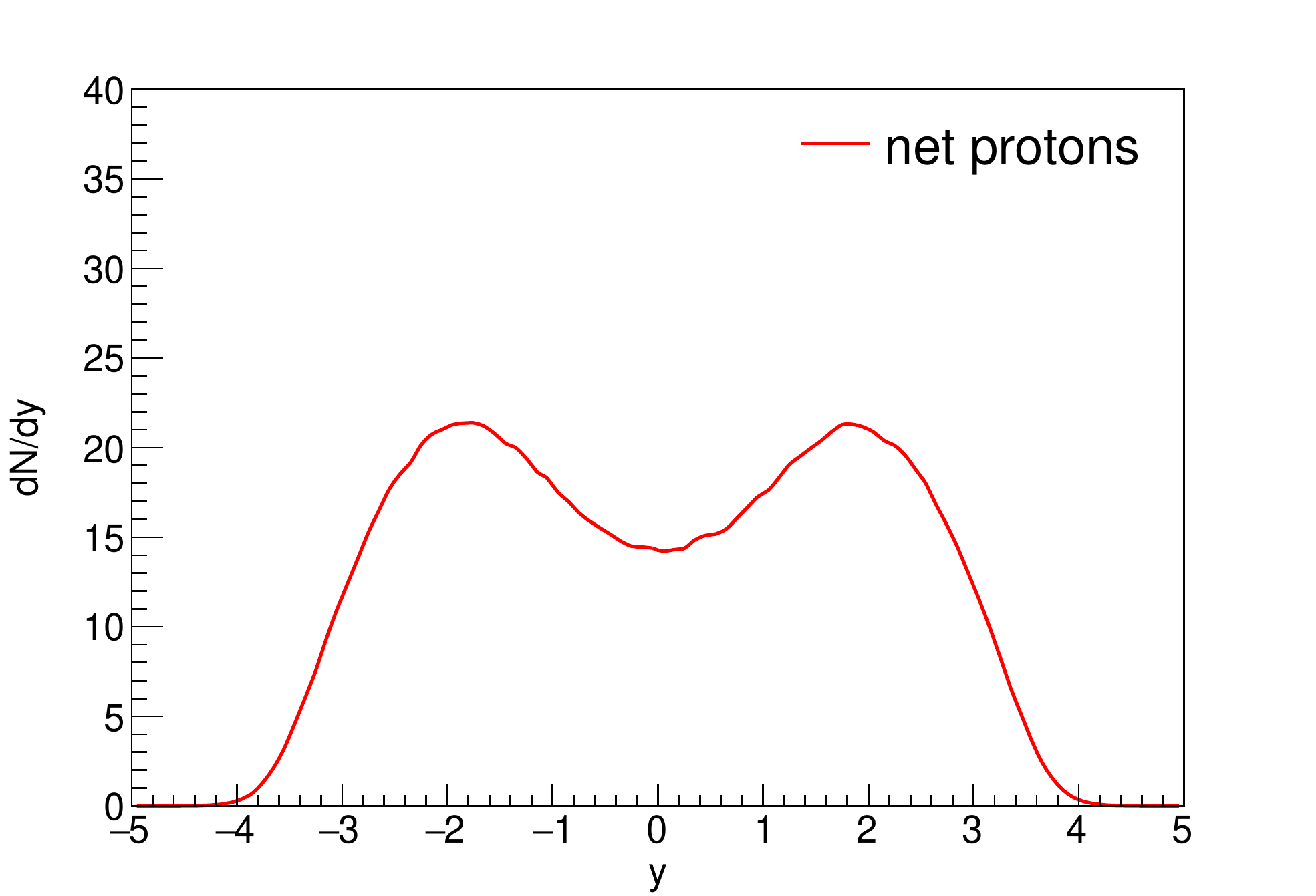}
 \caption{Rapidity distributions of identified hadrons (left panel) and rapidity distribution of net protons (right panel) at 0-5\% central Pb-Pb collisions at $\sqrt{s_{\rm NN}}=72$~GeV from the event-by-event model calculations.}\label{fig-dndy}
\end{figure}

The resulting rapidity distributions of pions, kaons and protons in the model are shown on Fig.~\ref{fig-dndy}. As the temperature at the particlization hypersurface is virtually flat up to geometrical rapidity $\eta=\pm3$, the rapidity shape of the pion and kaon yields is mostly defined by the change of the rapidity-differential effective volume $\Delta V_{\rm eff}=\int d\sigma_\mu u^\mu$ with geometrical rapidity. In addition to that, the increase of mean baryon chemical potential away from midrapidity enhances the production of baryons and suppresses antibatyons.
A shallow doubly-peaked structure in rapidity distribution for protons, combined with monotonic decrease of antiproton production with rapidity, results in somewhat more pronounced doubly-peaked structure for net protons, see right panel of Fig.~\ref{fig-dndy}. Comparing to existing results at $\sqrt{s_{\rm NN}}=62.4$~GeV from BRAHMS experiment at RHIC \cite{Arsene:2009aa} we conclude that the peak value of the net proton distribution is close to the experimental value, whereas the hole at midrapidity is not reproduced well due to the baryon charge distribution from UrQMD initial state.

\begin{figure}
 \begin{center}
 \includegraphics[width=0.6\textwidth]{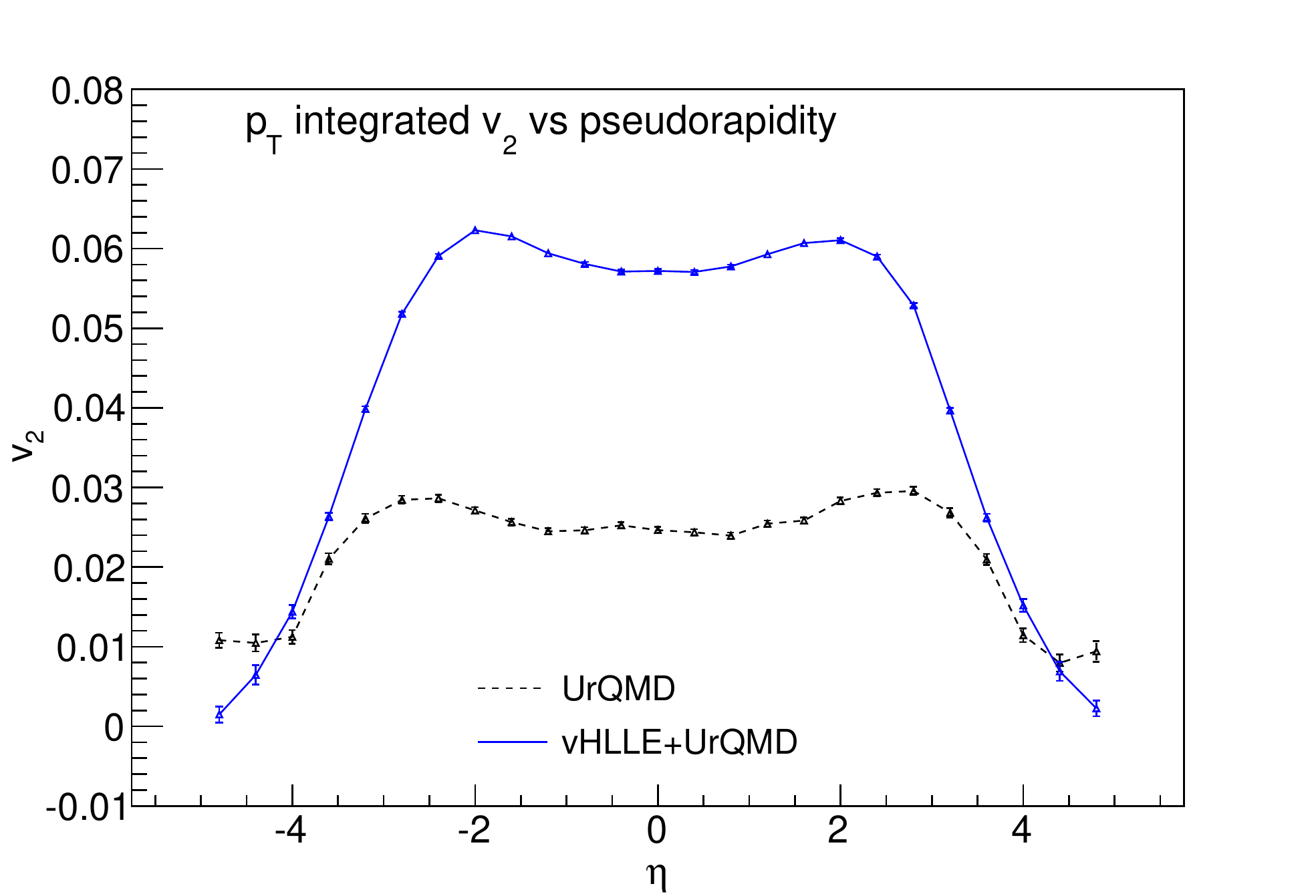}
 \end{center}
 \caption{Elliptic flow of all charged hadrons as a function of rapidity, for 20-30\% central Pb-Pb collisions at $\sqrt{s_{\rm NN}}=72$~GeV from the event-by-event model calculations.}\label{fig-v2}
\end{figure}

\begin{figure}
 \includegraphics[width=0.55\textwidth]{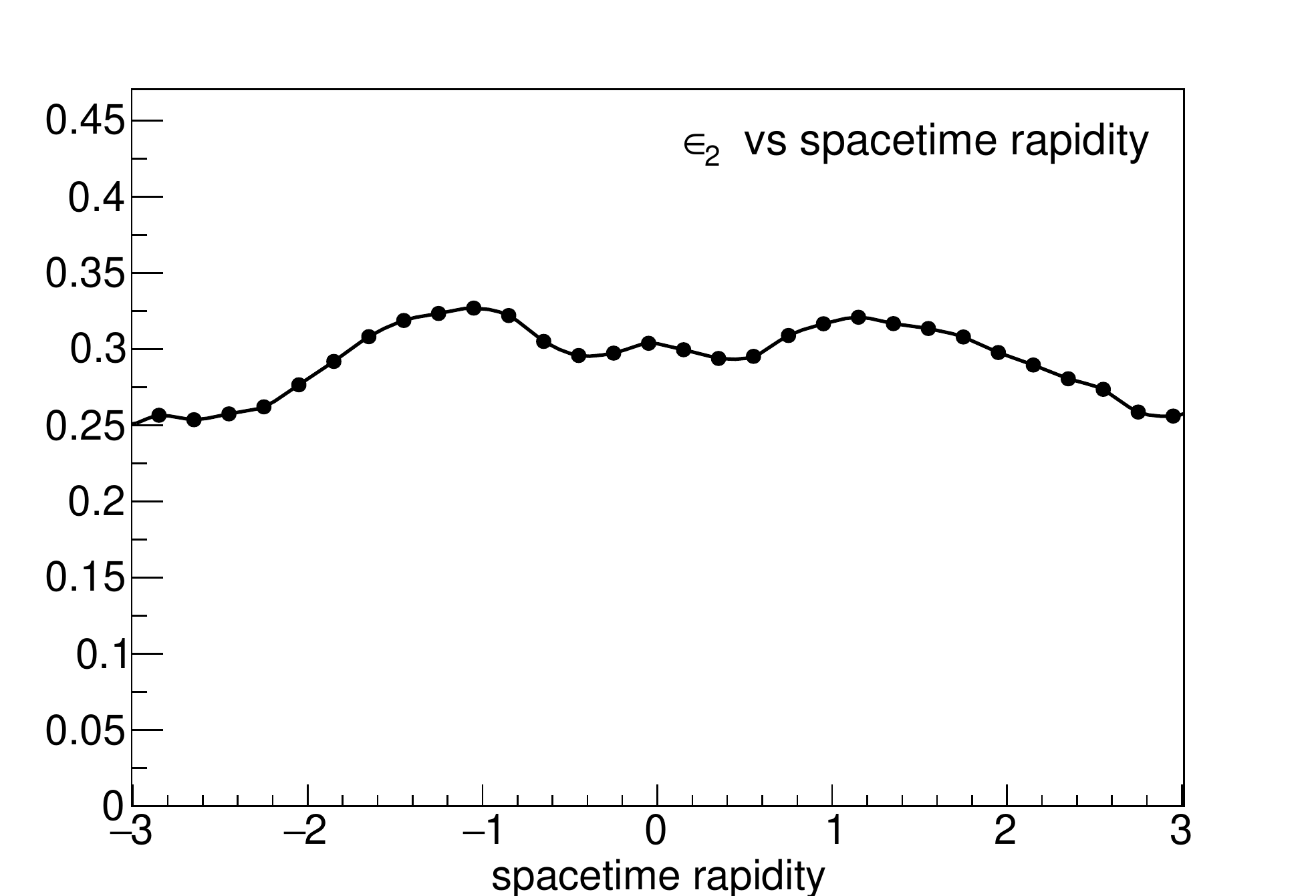}
 \includegraphics[width=0.55\textwidth]{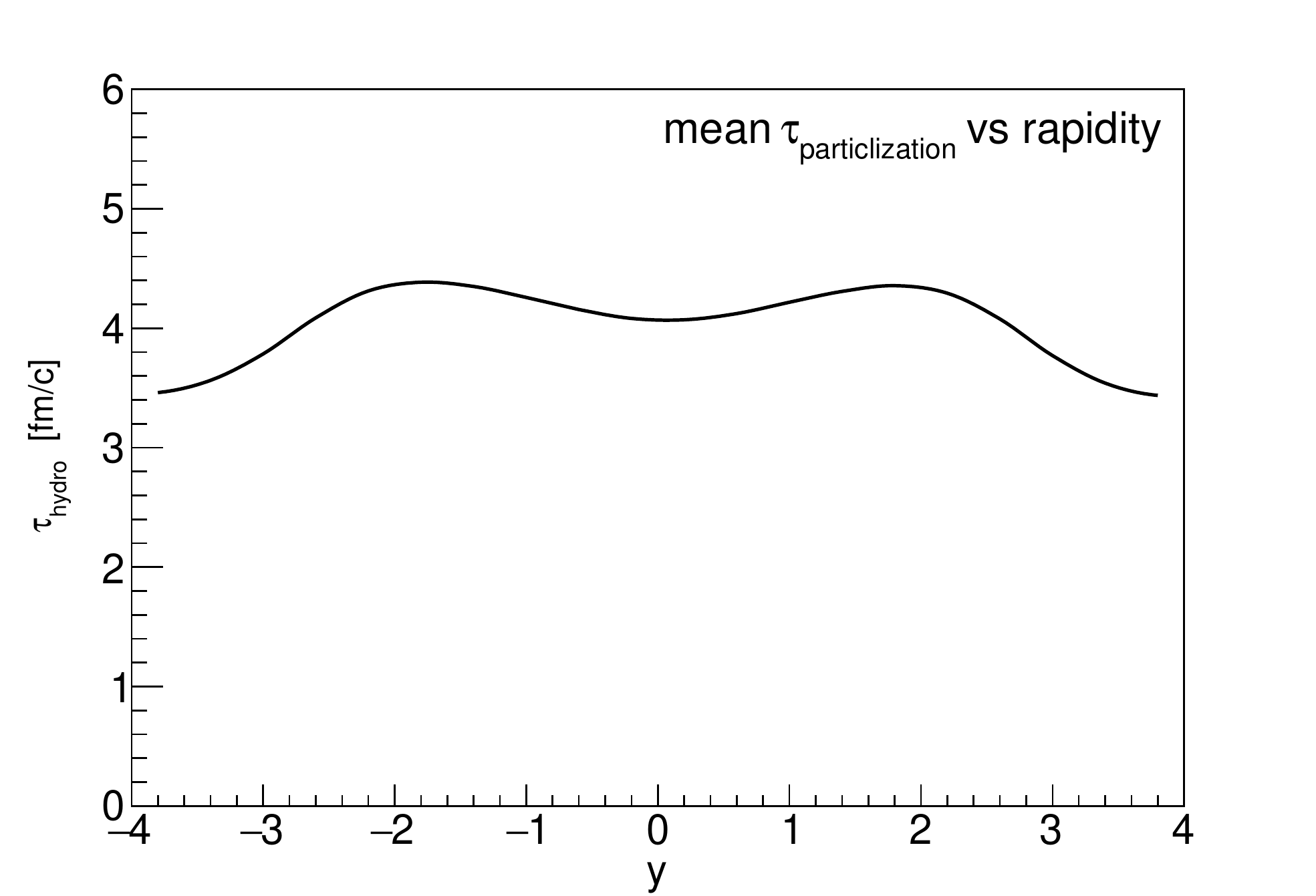}
 \caption{Eccentricity of initial state as a function of spacetime rapidity (left panel) and mean proper time of particlization as a function of momentum rapidity of particle (right panel) for 20-30\% central Pb-Pb collisions at $\sqrt{s_{\rm NN}}=72$~GeV from the event-by-event model calculations.}\label{fig-hydro}
\end{figure}

Finally we calculate the elliptic flow coefficient at different momentum rapidities using event plane method, and show the results on Fig.~\ref{fig-v2}. The elliptic flow exhibits peaks around pseudorapidity $\eta=\pm 2$. Such rapidity structure is driven by two effects: (i) the eccentricity of the initial state slightly increases as one departs from zero spacetime rapidity region of the initial state, see Fig.~\ref{fig-hydro}, left, and (ii) because of the doubly-peaked structure of the initial energy density the lifetime of the hydro stage as a function of geometrical rapidity peaks around $\eta=\pm2$.

\section{Conclusions}
In this work we have discussed the rapidity dependence of thermodynamic quantities and basic hadronic observables in $\sqrt{s_{\rm NN}}=72$~GeV PbPb collisions in the framework of viscous hydro+cascade model vHLLE+UrQMD. The initial state from the UrQMD cascade provides baryon, electric charge and strangeness densities along with the energy density. The initial baryon density profile is translated to the growth of effective baryon chemical potential with rapidity at the particlization (chemical freezeout), with no extra parameters.

More generally, different rapidity slices of the system cover different regions in the $T-\mu_{\rm B}$ plane during their evolution. Such a ``rapidity scan'' in addition to the collision energy scan can be a finer tool to study the QCD phase diagram at the RHIC BES energies.

\subsection{Acknowledgements}
The author acknowledges enligthening discussions with Dr.~Daniel Kiko\l{}a and Dr.~Barbara Trzeciak. The work is supported by Region Pays de la Loire (France) under contract no.~2015-08473.

%uncomment the following lines to place a figure
%\begin{figure}[htb]
%\centerline{%
%\includegraphics[width=12.5cm]{Fig1}}
%\caption{Plot of ...}
%\label{Fig:F2H}
%\end{figure}

\end{document}